\documentclass[reprint,amsmath,amssymb,aip,onecolumn]{revtex4-2}
\usepackage{xcolor}
\usepackage{acronym}
\usepackage{graphicx}
\usepackage[hidelinks]{hyperref}
\usepackage{booktabs}
\usepackage{tabularx}
\usepackage{array}
\usepackage{amsmath}

\newcolumntype{Y}{>{\centering\arraybackslash}X}
\acrodef{ai}[AI]{artificial intelligence}
\acrodef{ml}[ML]{machine learning}
\acrodef{pcm}[PCM]{phase-change memory}
\acrodef{rram}[RRAM]{resistive random-access memory}
\acrodef{imc}[IMC]{in-memory computing}
\acrodef{aimc}[AIMC]{analog in-memory computing}
\acrodef{mvm}[MVM]{matrix-vector multiplication}
\acrodef{dnn}[DNN]{deep neural network}
\acrodef{hbm}[HBM]{high-bandwidth memory}
\acrodef{dram}[DRAM]{dynamic random-access memory}
\acrodef{sram}[SRAM]{static random-access memory}
\acrodef{llm}[LLM]{large language model}
\acrodef{dl}[DL]{deep learning}
\acrodef{dlrm}[DLRM]{deep learning recommendation model}
\acrodef{adc}[ADC]{analog-to-digital converter}
\acrodef{snr}[SNR]{signal-to-noise ratio}
\acrodef{tpr}[TPR]{tensor-product representation}
\acrodef{nvm}[NVM]{non-volatile memory}
\acrodef{ca}[CA]{cellular automata}
\acrodef{auc}[AUC]{area under the curve}
\acrodef{pwm}[PWM]{pulse-width modulation}
\acrodef{dac}[DAC]{digital-to-analog converter}
\acrodef{kge}[kGE]{kilo gate equivalent}
\acrodef{fpga}[FPGA]{field-programmable gate array}
\acrodef{ha}[HA]{hardware-aware}
\acrodef{scr}[SCR]{superposed composite representation}

\newcommand{\vi}{\mathbf{v}_\text{i}}
\newcommand{\vj}{\mathbf{v}_\text{j}}

\newcommand{\vhatj}{\hat{\mathbf{v}}_\text{j}}
\newcommand{\vhati}{\hat{\mathbf{v}}_\text{i}}

\newcommand{\ki}{\mathbf{k}_\text{i}}
\newcommand{\kj}{\mathbf{k}_\text{j}}

\newcommand{\Sp}{\mathbf{S}}
\newcommand{\Mi}{\mathbf{M}_\text{i}}
\newcommand{\Snew}{\mathbf{S_\text{new}}}

\newcommand{\Nvecperarr}{N_\text{vec/arr}}
\newcommand{\Narr}{N_\text{arr}}
\newcommand{\Nvec}{N_\text{vec}}

\newcommand{\etpr}{\mathbf{\epsilon}_\text{tpr}}

\begin{document}
\title{Superposed analog memory for vector storage in neural networks}
\author{Athanasios Vasilopoulos} \email{atv@zurich.ibm.com} \affiliation{IBM Research Europe, Zurich Research Laboratory, 8803 R\"{u}schlikon, Switzerland} \affiliation{Department of Electrical and Computer Engineering, University of Patras, 26504 Patras, Greece}
\author{Patrick Diener} \affiliation{IBM Research Europe, Zurich Research Laboratory, 8803 R\"{u}schlikon, Switzerland}
\author{Michael Hersche} \affiliation{IBM Research Europe, Zurich Research Laboratory, 8803 R\"{u}schlikon, Switzerland}
\author{Abbas Rahimi} \affiliation{IBM Research Europe, Zurich Research Laboratory, 8803 R\"{u}schlikon, Switzerland}
\author{Abu Sebastian} \email{ase@zurich.ibm.com}\affiliation{IBM Research Europe, Zurich Research Laboratory, 8803 R\"{u}schlikon, Switzerland}\affiliation{Kirchhoff-Institut für Physik, Heidelberg University, 69120 Heidelberg, Germany}
\date{\today}

\begin{abstract}

The growing memory demands of artificial neural networks pose a major challenge for energy-efficient computing, particularly for models that rely on large collections of high-dimensional weight and embedding vectors. Conventional memory architectures store these vectors individually, requiring many memory cells per vector and limiting achievable storage density. Here we demonstrate an analog memory architecture that encodes multiple vectors in a distributed superposition within a single memory array. Individual vectors are recovered through in-memory matrix–vector multiplication, allowing the degree of superposition to be tuned to balance storage density against recall fidelity. This approach introduces a new scaling dimension for analog memory, in which effective storage density can be increased without a corresponding increase in the number of physical memory cells. The distributed representation also provides resilience to device-level noise. We implement the architecture on a prototype phase-change memory chip and experimentally evaluate it using embedding tables from deep learning recommendation models and large language models. The results demonstrate inference accuracy comparable to full-precision software baselines, establishing analog superposition as a promising route towards dense, energy-efficient memory for artificial intelligence.

\end{abstract}
\maketitle

%-----------------------------------------------------
% INTRODUCTION
%-----------------------------------------------------
\section{Introduction}\label{sec:introduction}

The rapid progress of \ac{dl} over the past decade has been driven not only by advances in algorithms, but mainly by advances in computing hardware, particularly specialized accelerators~\cite{Jouppi2017,Sze2017,Reuther2022,silvano2025ACMCompSurvey}. During inference, these \ac{ai} accelerators repeatedly access large sets of \ac{dnn} parameters stored in memory and deliver them to processing units for computation. Because the energy required to access memory often exceeds that required to perform arithmetic operations, the storage and movement of model parameters have emerged as major bottlenecks in \ac{ai} hardware. As neural networks continue to grow, increasing the amount of model information that can be stored close to the processing units has become increasingly important.

Most contemporary \ac{ai} accelerators rely on a memory hierarchy comprising \ac{sram} as high-performance cache memory on the processor chip and off-chip \ac{dram} as high-capacity system memory. However, the storage density of \ac{sram}, which typically requires a six-transistor cell, remains limited, and improvements through conventional semiconductor scaling have not kept pace with the rapid growth of modern neural networks. Embedded flash memory has traditionally served as on-chip storage for low-end edge platforms, but it becomes impractical to integrate at advanced technology nodes. Emerging memristive devices such as \ac{rram} and \ac{pcm} are expected to contribute significantly to integrated on-chip memory because of their compact size~\cite{Y2013yangNanoscale,lanza2022Science,hY2024hellenbrandMRS,Lanza2025}; their ability to represent data through binary high- and low-resistance states has already established a niche in edge-\ac{ai} processors for storing neural-network parameters~\cite{Y2025prabhuJSSC}. Nevertheless, conventional approaches to increasing memory capacity, namely, reducing device dimensions, enlarging arrays, or extending them into three dimensions, ultimately depend on the number of physical devices and the number of reliably distinguishable states per device. Moreover, depending on the memory technology, device scaling can increase device resistance, while larger arrays introduce greater interconnect losses, both leading to more demanding sensing conditions. Finally, although emerging non-volatile memory devices are intrinsically analog, their continuous state space is often restricted to a limited number of discrete levels to manage variability and sensing constraints.

Here we introduce an analog memory architecture that increases effective storage density through the representation of information rather than through physical device scaling alone. The proposed architecture adds a capacity-scaling dimension that arises from how information is represented, improves the sensing conditions that constrain memory-array scalability, and exploits the analog characteristics of emerging memory devices. Multiple high-dimensional vectors are stored across a fixed set of devices, and the number of vectors can exceed the number of physical rows of the array. This partially decouples effective storage capacity from the number of physical memory units. Increasing the superposition factor is accompanied by a gradual reduction in retrieval fidelity, providing a continuous density-fidelity trade-off that can be selected according to the requirements of the application. The approach is particularly suited to \ac{dnn} parameters, whose information is distributed across high-dimensional vectors and is often tolerant to reduced precision and moderate perturbations~\cite{gupta2015ICML,Rasch2023,Buechel2025a}.

The proposed memory architecture also changes the physical readout mechanism. Rather than sensing an individual memory device, a vector is retrieved by collectively activating the array rows and sensing the resulting column currents generated by many devices. Under the operating conditions examined here, this collective readout produces a median current more than two orders of magnitude larger than single-device readout, while reducing the coefficient of variation by more than sixfold. The resulting sensing margin can be traded against individual-device conductance, creating headroom for operation with more resistive nanoscale devices and extending the design space for larger memory arrays.

We demonstrate the operating principles and experimental realization of this architecture using a prototype \ac{pcm} chip. We establish the relationship between superposition factor and retrieval fidelity, quantify the sensing advantages of collective readout, and show how embedding tables from a \ac{dlrm} and an \ac{llm} can be adapted to superposed storage. Experimental retrieval from the \ac{pcm} chip preserves downstream inference performance comparable to full-precision software baselines, demonstrating the potential of analog superposition as a complementary scaling strategy for dense memory in \ac{ai} hardware.

%-----------------------------------------------------------
% THE CONCEPT
%-----------------------------------------------------------
\section{In-Depth Look at the Memory Architecture}\label{sec:memory_architecture}

\begin{figure}[ht]
    \centering
    \includegraphics[width=\textwidth]{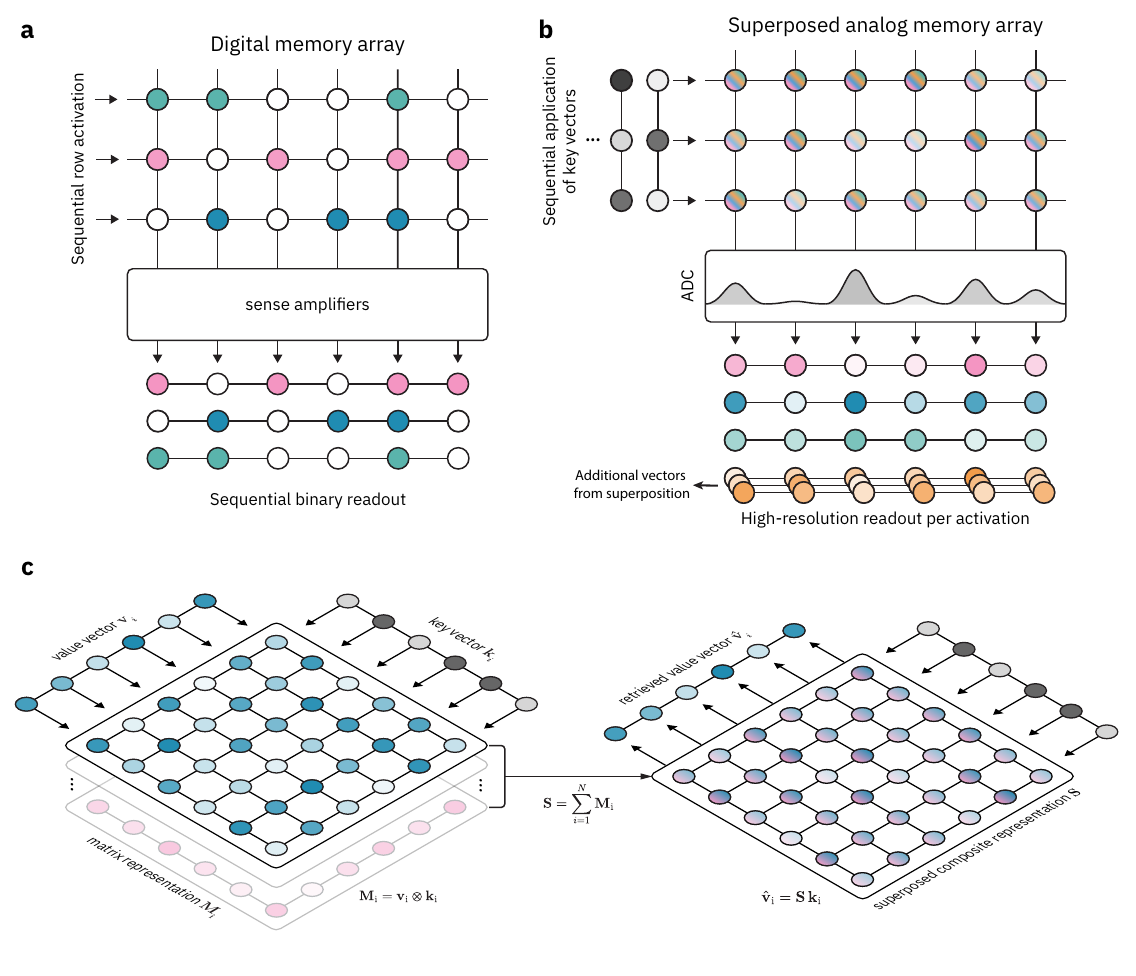}
    \caption{\textbf{From localized binary to distributed superposed analog storage.}
    \textbf{a,} Conventional digital memories store vector elements as binary states in individual cells and retrieve them through sequential row-by-row activation. The number of vectors that can be stored is therefore limited by the number of physical rows in the array. Colored circles denote logic \texttt{1}, while blank circles denote logic \texttt{0}. Different colors correspond to different vectors.
    \textbf{b,} In the proposed architecture, vector elements are stored as distributed analog states across the entire array and retrieved through collective activation of multiple rows. This collective readout increases the effective signal, and the encoding allows multiple vectors to be stored in superposition within the same set of devices, enabling storage capacity beyond the number of physical rows.
    \textbf{c,} Encoding and retrieval based on tensor-product representations. 
    Each value vector $\vi$ is bound with an associated key vector $\ki$ through an outer product to form a distributed matrix representation $\Mi$. Multiple bindings are then superposed to form the \ac{scr} $\Sp$. Individual vectors are retrieved by multiplying $\Sp$ with the corresponding key vector $\ki$, producing an approximate reconstruction of the stored value $\vhati$.
    }
    \label{fig:analog_memory_and_superposition}
\end{figure}

In conventional memory architectures, the elements of a \ac{dnn} vector are digitized, stored across individually addressable memory cells, and retrieved through sequential word or row accesses (Fig.~\ref{fig:analog_memory_and_superposition}a). The number of stored vectors is therefore determined by the number of available physical rows and the number of reliably distinguishable states per device. In the proposed architecture, each vector is instead distributed across an entire analog memory array and retrieved by collectively activating its rows according to an associated key (Fig.~\ref{fig:analog_memory_and_superposition}b). This representation allows multiple vectors to be superposed within the same devices, including operation where the number of stored vectors exceeds the number of physical rows. Furthermore, retrieval produces column currents from the collective response of many devices rather than from an individually addressed device, increasing the available current margin.

\subsection{Encoding scheme}
Prior to being written into the memory array, each $D$-dimensional vector is mapped to a two-dimensional representation that can be stored in a $D\times D$ memory array. To achieve this, we lift the vectors into a matrix representation and combine them into a single \ac{scr}. To achieve this, we employ \acp{tpr}~\cite{Smolensky1990}, a mathematical framework which defines the \textit{binding} of two vectors, a key and a value, using a tensor (outer) product operation. The result is a higher-order representation where information about both components is preserved in a distributed fashion. In our setting, the value corresponds to a \ac{dnn} vector that we wish to store in memory, while the key is an auxiliary vector that indexes the value within the superposed representation. The overall encoding and retrieval process is illustrated in Fig.~\ref{fig:analog_memory_and_superposition}c.

The first step is to bind each value vector $\vi$ with its corresponding key vector $\ki$, both of dimensionality $D$, producing a matrix representation $\Mi$:
\begin{equation}\label{eq:bind}
\Mi = \vi \otimes \ki = \vi \ki ^\top, \qquad \vi, \ki \in \mathbb{R}^D, \qquad \Mi \in \mathbb{R}^{D \times D}
\end{equation}

Multiple such bindings can then be superposed to form a single \ac{scr} $\Sp$ that contains information about all stored vectors.
\begin{equation}\label{eq:superposition}
\Sp = \sum_{i=1}^N \Mi = \sum_{i=1}^N \vi \ki ^\top, \quad \Sp \in \mathbb{R}^{D \times D}
\end{equation}
The superposition can be performed through direct addition, as in Eq.~\ref{eq:superposition}, or through update methods that improve representational capacity, such as the delta rule~\cite{Schlag2021}, described in the Methods.

This representation can then be stored directly in the memory array. Retrieval of a stored vector is performed by multiplying the composite representation with the corresponding key vector:
\begin{equation}\label{eq:unbind}
\vhatj = \Sp \, \kj = \left( \sum_{i=1}^N \vi \ki^\top \right) \kj = \sum_{i=1}^N \vi \, (\ki^\top \kj)=
\begin{cases}
\vj, & \text{if } \ki^\top \kj = 0 \ \forall i \neq j, \\
\vj + \sum_{i \ne j} \vi \, (\ki^\top \kj), & \text{otherwise} \\
\end{cases}
\end{equation}

The accuracy of the retrieved vector $\vhatj$ depends on the relationships among the key vectors. If the keys are orthogonal ($\ki^\top\kj=0$ for $i\neq j$), the original vector $\vj$ can be reconstructed exactly, which is possible only when the number of stored vectors satisfies $N\le D$. However, exact orthogonality is not required in practice for \acp{dnn}. When the keys are drawn randomly in a high-dimensional space, they tend to be approximately orthogonal -- a property sometimes referred to as the \textit{blessing of dimensionality}~\cite{Kanerva2009}. In this regime, the inner products $\ki^\top\kj$ remain small but non-zero, resulting in an approximate reconstruction $\vj + \etpr$, where $\etpr$ represents an interference term arising from the other stored vectors. Although this interference grows as more vectors are superposed, the direction of the retrieved vector remains largely preserved (see Extended Data Fig.~\ref{ext_fig:tpr_graceful_degradation}). This phenomenon, known as \textit{graceful degradation}~\cite{Smolensky1990}, enables a controllable trade-off between storage capacity and retrieval accuracy.

Crucially, this property allows the number of stored vectors $N$ to exceed the dimensionality $D$ of the memory array. We define the ratio \(S_{\mathrm{p}}=N/D\) as the \textit{superposition factor}. In this work, we intentionally exploit quasi-orthogonal keys to unlock this regime, enabling the superposed memory to store more vectors than the number of physical rows while maintaining the retrieval fidelity required for the downstream tasks.

\subsection{Physical storage and collective readout}\label{sec:read}

After its creation, the \ac{scr} \(\Sp\) is mapped directly onto a \(D\times D\) analog memory array, where each memory cell, comprising one or more analog devices, is programmed to represent the corresponding element of \(\Sp\). Programming can be performed using iterative read-write-verify techniques commonly employed in analog memory devices~\cite{Papandreou2011}. In contrast to digital parameter storage, the continuously programmable device conductances are used directly to represent the elements of the composite matrix. In this work, we only investigate the storage of parameters that are static during inference and as such \(\Sp\) does not require updating. However, one could incrementally update \(\Sp\) by leveraging the crystallization kinetics and the corresponding accumulative behavior of PCM devices~\cite{Karunaratne2022}.

Retrieval of vector \(\vhati\) corresponds physically to an in-memory \ac{mvm} between the stored matrix \(\Sp\) and the key \(\ki\) (Eq.~\ref{eq:unbind}). Keys need only be quasi-orthogonal between them, a condition that can be achieved with bipolar values. Therefore, the inputs to the \ac{mvm} is bipolar voltages representing the elements of \(\ki\) and are applied simultaneously to the array rows, with the resulting column currents being digitized to obtain the elements of \(\vhati\). This differs from conventional cell-addressed storage, in which the sensing conditions are determined primarily by the current of an individual device.

Collective readout provides a larger current margin while preserving the analog information encoded in the array. In the experimental conditions considered here, \ac{scr} retrieval produces a median current more than two orders of magnitude larger than single-device readout and reduces the coefficient of variation of the read-current distribution by more than sixfold (Extended Data Fig.~\ref{ext_fig:current}). This additional margin allows the use of more resistive nanoscale devices, without imposing strict sensing requirements that increase sensing time or the complexity of sensing circuitry as to not pollute a very faint signal.

\subsection{Organization of a tiled memory system}\label{sec:mem_system}

\begin{figure}[ht]
    \centering
    \includegraphics[width=\textwidth]{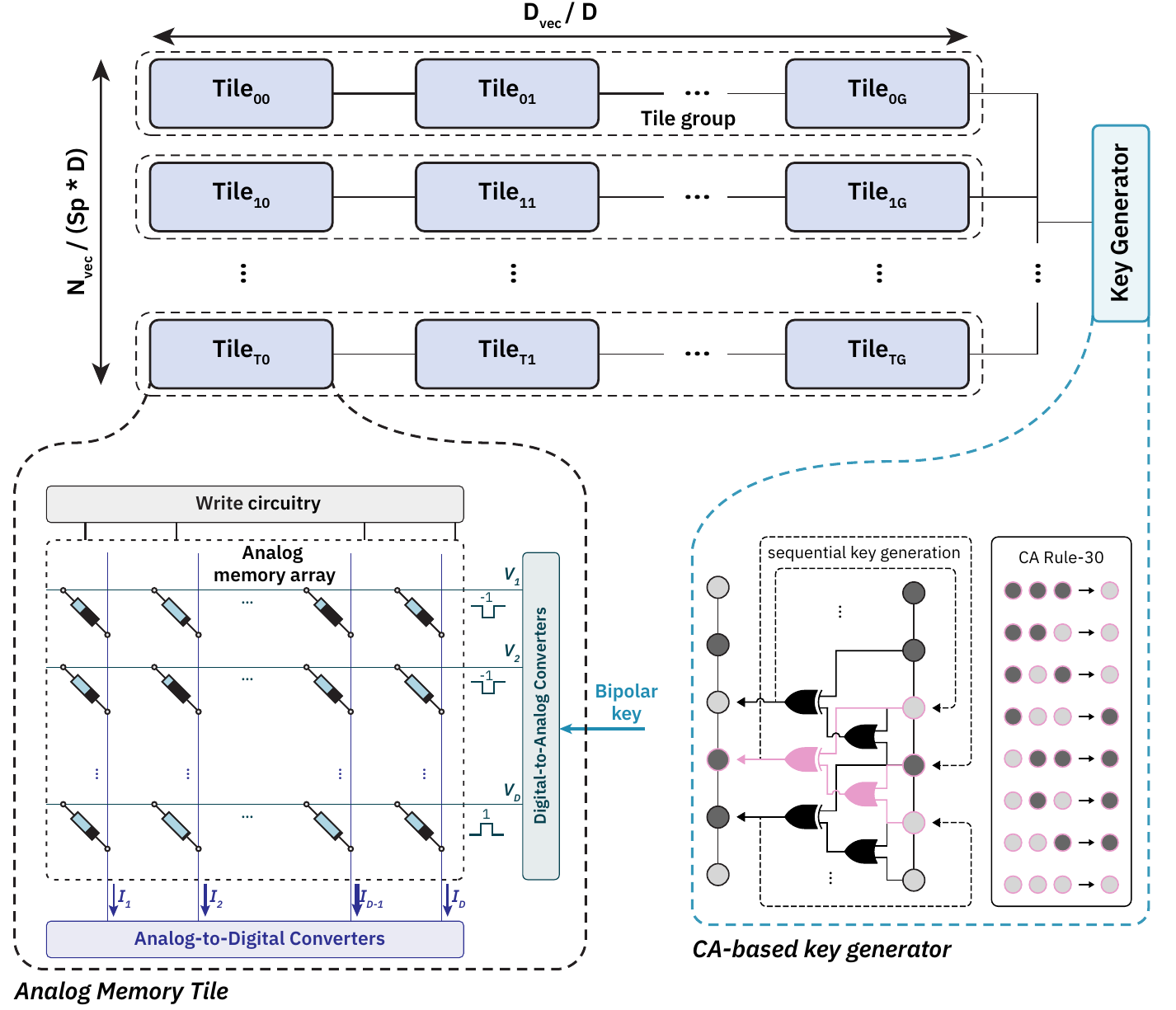}
    \caption{\textbf{Possible hardware realization and tiled organization of the proposed memory architecture.} 
    An individual analog memory tile stores one \ac{scr} $\Sp$ in a $D\times D$ crossbar, with each memory cell comprising one or more programmable analog devices that encode an element of $\Sp$. During retrieval, the key vector $\ki$ is converted to row voltages through a bipolar \acp{dac} and applied to the array, performing an in-memory matrix–vector multiplication. The resulting column currents are digitized by multi-bit \acp{adc} to produce the retrieved vector segment $\vhati$. Programming of the devices is performed through standard read-write verify circuitry. For vectors of dimensionality $D_{\mathrm{vec}}>D$, multiple tiles are organized into a tile group and operated in parallel, with each tile storing and retrieving one $D$-dimensional segment of the vector. The number of tiles per group is $\lceil D_{\mathrm{vec}}/D\rceil$, while a collection of $N_{\mathrm{vec}}$ vectors requires $\lceil N_{\mathrm{vec}}/(S_{\mathrm{p}}D)\rceil$ tile groups, where $S_{\mathrm{p}}=N/D$ is the superposition factor. A shared lightweight \ac{ca}-based key generator produces quasi-orthogonal bipolar keys using Rule~30 and distributes the selected key to the active tiles. Reusing the same keys across tile groups and the tiles within them amortizes key generation over many arrays and avoids explicit storage of the key vectors.
    }
    \label{fig:hardware_realization}
\end{figure}

Figure~\ref{fig:hardware_realization} illustrates how the proposed architecture can be organized using multiple analog memory tiles. Each tile contains a \(D\times D\) analog array, row \acp{dac}, column \acp{adc}, and the programming circuitry required to program a \ac{scr}. Vectors whose dimensionality \(D_{\mathrm{vec}}\) exceeds the tile dimension \(D\) are partitioned across $\lceil D_{\mathrm{vec}}/D\rceil$ tiles. These tiles form a tile group and are activated in parallel to retrieve the different segments of the same vector.

Within each tile group, one \ac{scr} stores \(N=S_{\mathrm{p}}D\) vectors. A collection of parameters containing \(N_{\mathrm{vec}}\) vectors therefore requires $\lceil N_{\mathrm{vec}}/{S_{\mathrm{p}}D}\rceil$ tile groups. Increasing \(S_{\mathrm{p}}\) reduces the required number of tile groups without changing the physical dimensions of the individual arrays, at the cost of the gradual retrieval-fidelity reduction introduced by superposition.

The same set of keys is reused across tile groups because the key identifies the position of a vector within a group, while the group address identifies the physical tiles containing that vector. Consequently, the key-generation circuitry does not need to be replicated for every analog tile. Instead, a shared generator supplies the selected bipolar key to the row-driving circuits of all tiles activated for a read operation. Its hardware cost is therefore amortized across many arrays and across all vector segments participating in the retrieval.

Finally, the fact that the keys need only be quasi-orthogonal bipolar vectors further simplifies the key generator circuit significantly. While a detailed design-space exploration of the key generator circuit falls out of scope for this work, for our simulations we adopt a hardware-inexpensive solution of keys that are generated sequentially via binary operations of an internal state, thus not requiring the storage of a separate key matrix. In this work, we use a \ac{ca} based on Rule 30 to generate pseudo-random bipolar sequences (see related Methods section).

%---------------------------------------------------------------------
% NEURAL NETWORK ADAPTATION AND MEMORY READOUT PRECISION
%---------------------------------------------------------------------
\section{Neural Network Adaptation}\label{sec:tpr_training_and_perf}

The memory architecture described in Section~\ref{sec:memory_architecture} stores vectors approximately due to the interference term $\etpr$ introduced by the \ac{scr}. Therefore, it is essential to adapt the parameters of the \ac{dnn} model to tolerate this interference. We achieve this by introducing the \ac{tpr}-based encoding during model retraining by injecting the retrieval error caused by superposition into the model parameters during the forward pass. Backpropagation then updates the parameters such that they become robust to the errors introduced by the \ac{scr} transformation, allowing the model to converge to parameters amenable to noisy storage in superposition. A detailed description of the training procedure is provided in Methods and in Extended Data Fig.~\ref{ext_fig:training}.

To assess downstream performance when model parameters are stored in $\Sp$, we consider two representative application scenarios where storage capacity is critical and where the inherent superposition capability of the transformation becomes advantageous. First, we evaluate a \ac{dlrm}~\cite{Naumov2019}, which contains large embedding tables that are repeatedly accessed during inference. Second, we apply the method to store the embedding table of a \ac{llm}, specifically GPT2-large~\cite{radford2019language}. Details on the selected networks and datasets are provided in Methods, while the hyperparameters used during training are listed in Extended Data Table~\ref{ext_tab:hyperparams}.

\begin{figure}[ht]
    \centering
    \includegraphics[width=\textwidth]{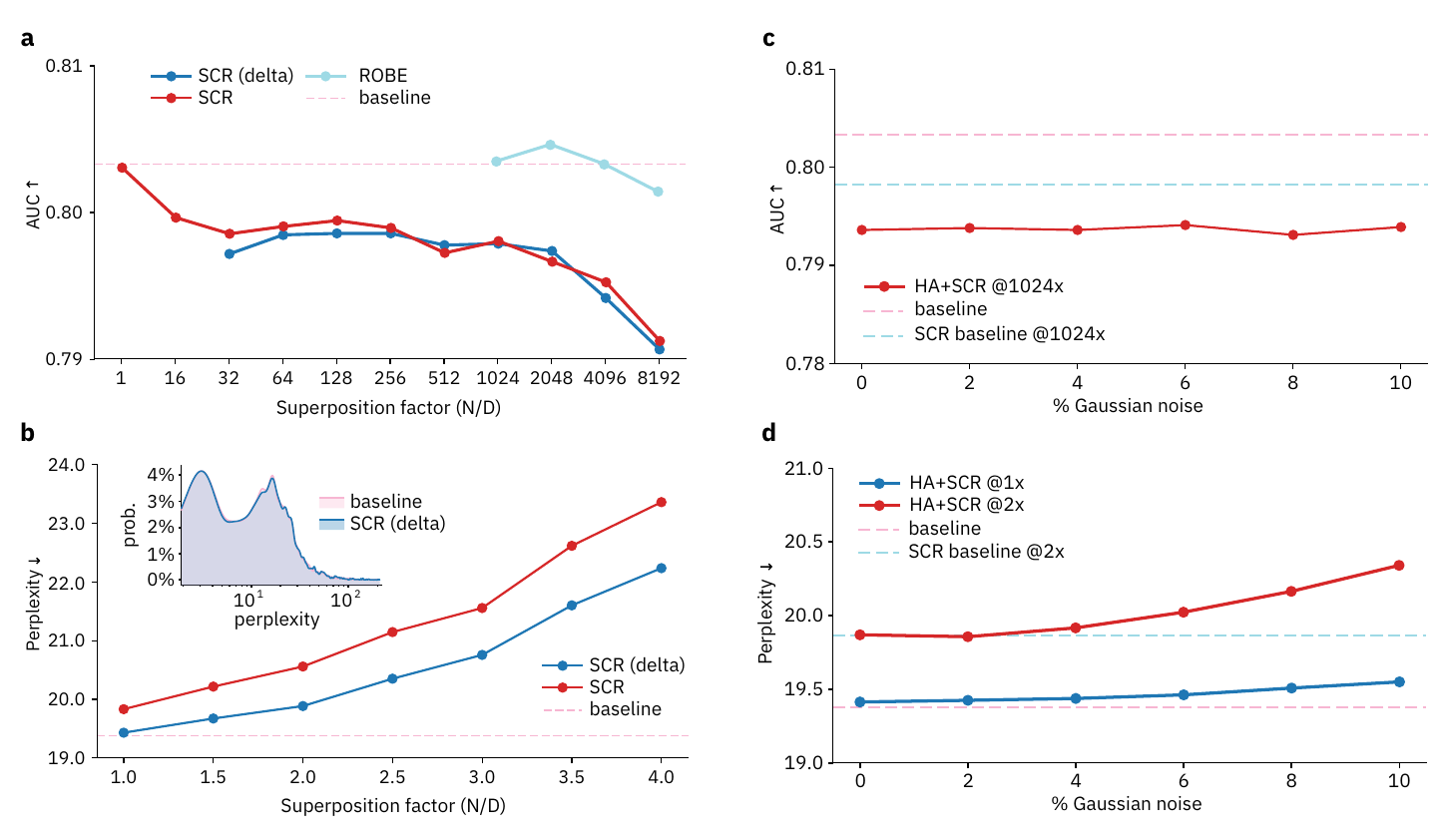}
    \caption{\textbf{\ac{scr}-aware adaptation of \ac{dnn} vectors.} 
    \textbf{a,} Performance of a \ac{dlrm} on the Criteo Kaggle Display Advertising dataset when embedding tables are stored in superposition using \acp{tpr}. Baseline accuracy is preserved when storing $N=D$ vectors (no compression), while increasing $N$ enables compression of the embedding table with only minor degradation in \ac{auc}. Performance remains within $0.01$ \ac{auc} of the baseline and of the state-of-the-art compression method ROBE. \textbf{b.} Perplexity of the GPT2-large on WikiText when the embedding table is stored in superposition. Baseline performance is recovered for $N=D$ (using the delta \ac{scr} variant) while increasing $N$ produces a graceful degradation in perplexity, enabling a tunable trade-off between compression and model accuracy. Inset: the perplexity distribution for the baseline model and for $N=D$ remains nearly identical.
    \textbf{c,} Robustness of \ac{dlrm} embeddings stored in superposition ($N=1024D$) under injected Gaussian noise representing analog hardware non-idealities. Networks trained with hardware-aware noise injection maintain iso-accuracy -- less than 1\% drop from baseline -- even for high noise levels. \textbf{d,} Noise robustness of GPT2-large embeddings stored in superposition for $N=D$ and $N=2D$. Language models exhibit greater sensitivity to noise, with modest perplexity degradation at higher noise levels and slightly increased sensitivity under compression. Arrows in the y-axis labels indicate whether higher ($\uparrow$) or lower ($\downarrow$) values correspond to better performance.
    }
    \label{fig:sim_results}
\end{figure}

We first analyze the impact of storing the embedding tables of the \ac{dlrm} using \ac{scr}, evaluated in the Criteo Kaggle Display Advertising dataset~\cite{criteo-display-ad-challenge} and measured in \ac{auc} (Fig.~\ref{fig:sim_results}a). When the embeddings are stored in memory arrays with superposition factor equal to one ($N=D$), the resulting \ac{auc} matches the software baseline, indicating that the encoding does not affect downstream performance. However, since embedding tables in \acp{dlrm} are known to be highly compressible~\cite{Shi2020}, we further investigate the behavior of the model as the number of stored vectors $N$ increases way beyond $D$. Figure~\ref{fig:sim_results}a shows that performance remains close to the software baseline even at superposition ratios as high as $1024\times$, with the model remaining within $0.01$ \ac{auc} of both the baseline and the state-of-the-art embedding compression method ROBE~\cite{Desai2022}. Notably, our approach introduces no decompression overhead, as retrieval from the \ac{scr} results to the target vector naturally.

We perform a similar analysis for GPT2-large by storing its embedding tables in $\Sp$ and measuring perplexity on WikiText~\cite{wikitext}. Here, on a network much more sensitive to interference, we observe that we need to employ the delta rule variant of \ac{scr} to achieve baseline perplexity when we store $N=D$ vectors. As superposition increases, perplexity degrades gracefully, reflecting the interference introduced by superposition and the tradeoff of storage density to retrieval fidelity. Importantly, storing embeddings using \ac{scr} without compressing ($N=D$) does not alter the perplexity landscape of the network compared to the baseline model, as shown in the inset of Fig.~\ref{fig:sim_results}b, denoting no observable change in the quality of the results.

From these studies, we observe that networks trained with \ac{scr} awareness can achieve baseline accuracy without compression on both tasks, while also providing a flexible capacity to retrieval fidelity tradeoff without any overhead cost during retrieval. Furthermore, the achievable superposition can be tuned continuously by adjusting the number of stored vectors $N$, rather than being limited to coarse compression factors such as $2\times$ or $4\times$, common in digital systems. Finally, the performance of the method generalizes across both model types as the encoding process is distribution-agnostic, as opposed to highly specialized compression methods developed for specific network architectures, like ROBE~\cite{Desai2022} (see Extended Data Fig.~\ref{ext_fig:gpt_robe}).

Besides \ac{scr} awareness, robustness to analog hardware non-idealities must also be ensured. Analog memory devices exhibit conductance fluctuations, and in-memory \ac{mvm} operations introduce additional effects such as IR-drop and output discretization~\cite{Singh2025,Rasch2023}. To account for these factors, we extend the training procedure to include analog \ac{ha} elements. Following standard practice, we introduce Gaussian noise, clipping, and \ac{adc} quantization during training to improve robustness to hardware imperfections~\cite{LeGallo2023,Rasch2023}. Details of the \ac{ha} training procedure are described in Methods. 

To verify robustness, we measure the performance of \ac{ha}-trained networks under varying levels of injected Gaussian noise in the elements of $\Sp$ during evaluation, as a proxy for the device noise~\cite{Rasch2023}. Figures~\ref{fig:sim_results}c,d show the results for \ac{dlrm} at $1024\times$ superposition and for GPT2-large at $1\times$ and $2\times$ superposition. The \ac{ha} simulations demonstrate that performance is sustained even in the presence of substantial noise -- up to $10\%$ variation in the stored representation.

%--------------------------------------------------------------------------
%  HARDWARE EXPERIMENTS
%--------------------------------------------------------------------------
\section{Experimental validation}\label{sec:hardware}
We validate the proposed memory architecture through hardware experiments using the \ac{pcm}‑based \ac{aimc} chip described in~\cite{LeGallo2023,Syed2023}. The chip consists of 64 memory tiles, each implementing a $256\times256$ crossbar array of unit cells based on \ac{pcm} devices. The chip -- fabricated in 14nm CMOS technology node -- integrates all required peripheral circuitry to program the \ac{pcm} devices to target analog conductance states and to carry out in‑memory \ac{mvm} operations. 

\begin{figure}[ht]
    \centering
    \includegraphics[width=\textwidth]{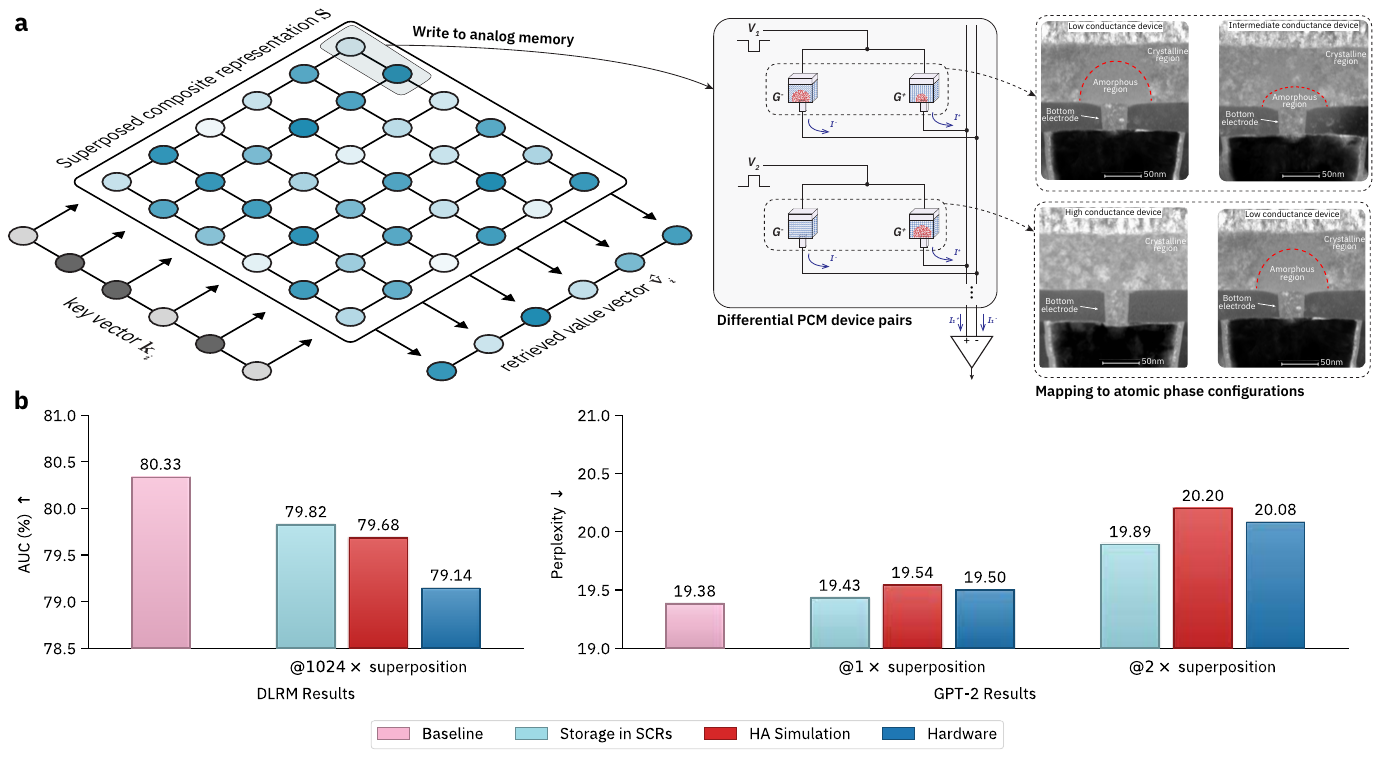}
    \caption{\textbf{Experimental validation on PCM hardware.}
    \textbf{a,} Physical storage of the superposed composite representation $\Sp$ in \ac{pcm} devices. Each element of $\Sp$ is encoded using a differential pair of \ac{pcm} devices, whose conductance difference represents the signed analog value of the corresponding matrix element. The conductance states are manifested as different amorphous–crystalline phase configurations within the \ac{pcm} material, enabling storage of analog values directly in the memory array.
    \textbf{b,} Hardware inference results when embedding tables are stored in superposition on the \ac{pcm}-based \ac{aimc} chip. Vectors are retrieved through in-memory matrix–vector multiplication and digitized on chip before being used for inference in software. Measured performance closely follows the hardware-aware simulation results and remains near the software baseline, with only modest degradation corresponding to effective noise levels of approximately $8$–$10\%$ in the stored representation. Arrows in the y-axis labels indicate whether higher ($\uparrow$) or lower ($\downarrow$) values correspond to better performance.
    }
    \label{fig:hw_results}
\end{figure}

To match the hardware tile dimensions, we construct a tiled representation of the network embeddings evaluated in Figures~\ref{fig:sim_results}c,d using $D=256$. The \ac{scr} $\Sp$ is then mapped directly onto the \ac{pcm} array. Each element of $\Sp$ is encoded using a differential pair of \ac{pcm} devices, whose conductance difference represents the signed analog value of the corresponding matrix element (Fig.~\ref{fig:hw_results}a). The conductance states are realized physically through different amorphous–crystalline phase configurations within the \ac{pcm} material, enabling the storage of analog values in the memory array.

Retrieval is performed directly on chip through in-memory \ac{mvm} operations with the key vectors $\{\ki\}$. During readout, the rows of the array are biased with $0.2\,\mathrm{V}$ read voltages according to the key vector values, producing column currents proportional to the values of the stored vectors $\{\vi\}$. These currents are digitized using current‑controlled oscillator‑based \acp{adc} and converted on‑chip into signed 8‑bit integers, which we used as the final readout values. More details about the experimental platform can be found in Methods.

These hardware‑generated parameters are subsequently used to perform inference in software to evaluate model accuracy. Figure~\ref{fig:hw_results}b summarizes the resulting performance across the evaluated networks. The measured accuracy closely matches the software baselines and aligns with the simulation results reported in the previous section for noise levels of approximately $8-10\%$. These results confirm that the proposed superposed analog storage scheme can be realized on existing \ac{pcm} hardware while preserving the accuracy benefits predicted by simulation.

%--------------------------------------------------------------------------
%  DISCUSSION AND CONCLUSION
%--------------------------------------------------------------------------
\section{Discussion}\label{sec:outlook}

% Performance advantages
The proposed memory architecture addresses several key challenges in the design of memory devices and large‑scale arrays. By replacing conventional row‑by‑row readout with collective readout through in‑memory \ac{mvm}, it relaxes the minimum‑conductance requirements for reliable sensing while simultaneously increasing the effective read signal. Because the readout current arises from multiple devices operating in unison rather than from a single memory cell, the architecture tolerates substantially higher device resistances. This, in turn, allows the use of smaller devices~\cite{Pi2018}, enabling further increases in array density and crossbar size. In addition, the collective read mechanism can deliver substantially faster read operations even when the underlying devices exhibit very high resistances. From a performance and energy efficiency standpoint, we provide array-level projections of latency, bandwidth, and energy per effective bit in Extended Data Table~\ref{ext_tab:performance_proj}. We do not attempt direct comparisons with \ac{sram} or \ac{dram} memory systems, as our projections correspond to a single array and do not include the full overhead of modern memory hierarchies. Nevertheless, the array-level performance and the key properties of the proposed memory architecture -- non-volatility, on-chip integration, and superposition without decompression overhead -- suggest a credible research path towards integrating many such arrays within future memory systems.

Moreover, the architecture provides a path for exploiting noisy analog device states. Although reliable multi-bit storage remains challenging at the level of individual \ac{pcm} devices, the relevant metric in the proposed architecture is the accuracy of the collective \ac{mvm} output. In Extended Data Fig.~\ref{ext_fig:mvm_accuracy}, we show that, in one of the cases explored in the paper, the retrieval error lies between that of 3- and 4-bit digital reference, closer on average to the 4-bit reference. This is consistent with the effective \ac{mvm} precision previously demonstrated for \ac{dnn} inference~\cite{Vasilopoulos2023}. This suggests that useful multilevel memory characteristics can also emerge from the collective response of devices whose individual states are noisy, opening a path towards exploiting scaled analog memories without requiring precise resolution of each device state.

% Dual use
Beyond improving on-chip storage capacity, the proposed architecture also opens opportunities for rethinking the role of \ac{aimc} hardware. In particular, \ac{aimc} tiles could be used not only for computation but also as high-density vector storage units with minimal hardware and control overhead. For example, in multi-tile \ac{aimc} chips~\cite{wan2022Nature, Ambrogio2023, Khwa2025}, some tiles could perform in-memory computation while others store \ac{dnn} vectors such as embeddings. Such a configuration could reduce the reliance of \ac{aimc} systems on external memory accesses~\cite{Jain2023} and may be particularly beneficial for cost- and energy-sensitive markets such as edge computing, where large off-chip \ac{dram} resources may be impractical~\cite{Boybat2024}.

% Opportunities for stand-alone memory, extensions to analog memory such as flash
Although the primary focus of this work is to address challenges related to on‑chip storage, the proposed memory architecture can also be applied to standalone analog memories, including Flash storage. In combination with emerging 3D analog memory technologies, this architecture could potentially enable the storage of trillions of \ac{dnn} vectors within an extremely compact area and with minimal energy consumption. Notably, 3D analog memory is already being investigated for use in in‑memory \ac{mvm} operations \cite{Buechel2025}, suggesting a natural extension of the proposed architecture, by mapping the depicted tile groups of Fig.~\ref{fig:hardware_realization} on separate tiers of the same tiles.

From an algorithmic perspective, we introduced an \ac{scr}- and analog-aware training procedure to adapt the network parameters. Although such retraining is common when deploying models on reduced-precision hardware -- including analog or low-bit quantized platforms -- it can become prohibitively expensive in the era of \acp{llm}~\cite{Buechel2025a,Liu2024}. Future work could therefore explore methods to reduce this overhead. For instance, a lightweight post-training adaptation step that compensates for the approximate recovery of embeddings could significantly reduce retraining overhead. Similarly, limiting parameter updates to a small subset of layers -- by freezing most of the network and updating only a few layers following the embedding stage -- may allow the remaining layers to remain unchanged while preserving overall performance.

In summary, we introduced a memory architecture that stores high-dimensional \ac{dnn} vectors as distributed analog superpositions, providing an additional scaling dimension beyond the number of physical devices and their individually distinguishable states. We demonstrated its realization using two-dimensional analog memory arrays with the required peripheral circuitry, developed training procedures that compensate for superposition and hardware non-idealities, and experimentally validated the storage and retrieval of embedding tables from a \ac{dlrm} and an \ac{llm} on a prototype \ac{pcm} chip. Beyond these application-level demonstrations, the architecture offers a new perspective on nanoscale memory design. By exploiting collective readout and distributed representations, it creates opportunities to use more resistive, noisy and densely integrated devices without requiring each cell to provide the precision or readout conditions demanded by conventional memory architectures. This could relax constraints associated with device miniaturization, multilevel-state reliability and array scaling, while enabling more effective use of the intrinsic analog characteristics of emerging memory technologies. More broadly, our results suggest that information representation can become a design variable alongside device materials, geometry and array organization, opening a route towards co-optimizing nanoscale devices and memory architectures for the storage demands of future neural-network hardware.

%--------------------------------------------------------------------------
%  METHODS
%--------------------------------------------------------------------------
\clearpage
\section*{Methods}
\renewcommand{\theequation}{M\arabic{equation}}
\setcounter{equation}{0}
\subsection*{Superposition using a modified delta update rule}\label{meth:delta}
An alternative way to superpose tensor product representations to the element-wise addition of Eq.~\ref{eq:superposition} is the method of delta update, adapted from the memory update rule introduced in linear transformers by Schlag et. al.~\cite{Schlag2021}. In an iterative process, a new key-value pair $(\mathbf{k}_\text{new}, \mathbf{v}_\text{new})$ is added to the existing superposed representation $\Sp$ by first unbinding $\Sp$ with the newly drawn key vector $\mathbf{k}_\text{new}$ to yield $\hat{\mathbf{v}}_\text{new}$. This vector contains the information already stored in $\Sp$ that would interfere with $\mathbf{v}_\text{new}$ if it were to be added as is.

\begin{equation}\label{meth_eq:unbind_existing}
\hat{\mathbf{v}}_\text{new} = \Sp \cdot \mathbf{k}_\text{new}
\end{equation}

A linear combination is then formed between the existing value vector $\hat{\mathbf{v}}_\text{new}$ and the new value vector $\mathbf{v}_\text{new}$, with a factor $\beta$ controlling the flow of new information in the representation. Note that in our modified rule, we keep $\beta$ as a data-independent hyperparameter, while in the formulation by Schlag et. al.~\cite{Schlag2021}, it is a function of the transformer token.

\begin{equation}\label{meth_eq:linear_comb}
\mathbf{v}_\text{delta} = \beta \cdot \hat{\mathbf{v}}_{\text{new}} + (1 - \beta) \cdot \mathbf{v}_\text{new}
\end{equation}

The convex combination of old and new information is subsequently bound with the new key $\mathbf{k}_\text{new}$ and added to the existing memory matrix $\Sp$. At the same time, we remove the contribution of the existing part $\hat{\mathbf{v}}_\text{new}$ as its participation is now represented in the convex combination $\mathbf{v}_\text{delta}$.

\begin{equation}\label{meth_eq:delta_update}
\Snew = \Sp + \underbrace{(\mathbf{v}_\text{delta} \otimes \mathbf{k}_\text{new})}_{\text{write}}- \underbrace{(\hat{\mathbf{v}}_\text{new} \otimes \mathbf{k}_{\text{new}})}_{\text{remove}}
\end{equation}

While prior work supports that this way of superposing increases the representational capacity of the memory $\Sp$, the process serializes the creation of the memory, as one vector is added at a time. To improve runtime, we further modify this method with a batched implementation of the delta update rule, where a complete batch of keys $\{ \mathbf{k}_\text{1}, \mathbf{k}_\text{2}, \dots, \mathbf{k}_\text{B} \}$ and their corresponding values $\{ \mathbf{v}_\text{1}, \mathbf{v}_\text{2}, \dots, \mathbf{v}_\text{B} \}$ are updated at the same time, using batched versions of the equations~\ref{meth_eq:unbind_existing}-~\ref{meth_eq:delta_update}. We use the batched implementation for our experiments, as using the serialized version of the algorithm during retraining adds significant overhead to the runtime.

\subsection*{Key generation}\label{meth:keygen}
We consider three methods to generate keys: one to generate orthogonal keys, and two to generate quasi-orthogonal keys, based on pseudo-random processes.

To generate the orthogonal keys used in the analysis of Extended Data Fig.~\ref{ext_fig:tpr_graceful_degradation}, we utilize Hadamard matrices. Hadamard matrices provide a deterministic and efficient way to generate orthogonal keys~\cite{Hedayat1978}. A Hadamard matrix $\mathbf{H} \in \{-1, +1\}^{D \times D}$ satisfies the equation $\mathbf{H} \mathbf{H}^\top = D \cdot \mathbf{I}$. Each row (or column) of $\mathbf{H}$ is orthogonal to all others, making them ideal choices for key vectors.

The quasi-orthogonal keys used in Extended Data Fig.~\ref{ext_fig:tpr_graceful_degradation} are sampled from the Rademacher distribution~\cite{Vershynin2018}, where each element is drawn independently from $\{-1, +1\}$ with equal probability. Given sufficiently large dimensionality $D$, these vectors are quasi-orthogonal~\cite{Kanerva2009}. 

A more hardware-friendly approach to implement random sampling to create quasi-orthogonal vectors is to use the \ac{ca} Rule 30~\cite{Wolfram1983}. Rule 30 is a one-dimensional binary cellular automaton known for producing pseudo-random sequences. Starting from a seed vector drawn from the Rademacher distribution, we evolve it using Rule 30 to generate successive key vectors. Each $\ki$ is then mapped from binary values to $\pm1$. The transition matrix of \ac{ca} Rule 30 is included in Fig.~\ref{fig:hardware_realization}. Due to its hardware-friendly nature, we use this process to generate the keys in all the neural network experiments presented in the paper.

\subsection*{SCR-aware training}\label{meth:tpr_training}
To enable \ac{dnn} parameter storage in the proposed memory system, the networks must be adapted to be robust to the interference $\etpr$ introduced by the \ac{scr}. To achieve this, we devise a training procedure that can be applied either during training from scratch or fine-tuning. During this procedure, we introduce the \ac{tpr} transformation during the forward pass of the model, adding on the parameters the error term $\etpr$. That way, the backpropagation modifies the parameters to find optima that both lead to good task performance and to robustness to the \ac{tpr} transformation. 

In Extended Data Fig.~\ref{ext_fig:training}, we demonstrate the steps to introduce \ac{tpr} in the forward pass. At a first step, the parameters of the model $\mathbf{P}$ that are to be stored in the \ac{scr} $\Sp$ are collected and mapped in a set of value vectors $\mathbf{V}$ of dimension $D$, which represents the size of the memory array in the hardware. To achieve this, the parameters are all flattened in a single one-dimensional vector and then reshaped in the shape $(\Nvec, D)$, where $\Nvec$ is the number of all value vectors necessary to store all parameters. To achieve an integer number $\Nvec$, we pad with zeros when necessary. Then, according to the selected superposition factor, i.e., the number of vectors stored per memory array $\Nvecperarr$ of dimension $D$, we tile the parameters in a shape $(\Narr, \Nvecperarr, D)$, where $\Narr$ is the number of memory arrays required to store all parameters $\mathbf{P}$. 

In the second step, we generate all the unique keys required to bind and unbind the value vectors. Note that we only require $\Nvecperarr$ keys, as the different memory arrays $\Narr$ can share the same keys. It is crucial that the keys generated here agree with the implementation found in hardware, as the keys used here must be used during the hardware experiments as well. For our experiments, the generation of these keys happens via the \ac{ca} Rule 30, which, as mentioned in the previous Methods section, is a hardware-friendly method to generate quasi-orthogonal keys.

In the third step, the value vectors $\mathbf{V}$ are bound with the key vectors $\mathbf{K}$ to produce the \ac{scr} $\Sp$. Notably, since the value vectors are of shape $(\Narr, \Nvecperarr, D)$, the \ac{scr} is in reality a tiled representation comprising $\Narr$ matrices of shape $(D,D)$. Then, on step 4, using the same keys $\mathbf{K}$, we retrieve all values $\hat{\mathbf{V}}$ from the \ac{scr} matrices via an unbinding operation and finally on step 5 we map them back to the original set of parameters $\hat{\mathbf{P}}$, reversing the tiling and padding of the first step.

Due to the inaccurate retrieval process of the transformation, the retrieved vectors contain an error term, which is isolated for each parameter $\etpr=\hat{\mathbf{P}} - \mathbf{P}$. Finally, in step 6, as in a conventional noise injection-based training procedures, we perturb the parameters of the model according to the collected error terms, such that during the forward pass the model uses $\hat{\mathbf{P}}$, while during the backwards pass the backpropagation updates $\mathbf{P}$. After an update step takes place (step 7), and a new set of parameters $\mathbf{P}$ is found, we repeat the bind and unbind process all over.

This procedure adds overhead compared to the conventional fine-tuning of the model. However, we vectorize all binding and unbinding operations using the Einstein summation notation (\texttt{einsum}) offered by the PyTorch library. This optimization leads to the overhead being on par with other noise-injection or quantization-aware training procedures, such as ones found in popular frameworks~\cite{LeGallo2023a,Vasilopoulos2025}. Using the delta update rule to achieve superposition, though, inflicts further overhead, as the construction of the memory matrices $\Sp$ happens in a serial manner. However, we overcome this by adopting a batched version of the delta update rule, which reduces the serialization overhead to an acceptable amount, trading off the effectiveness of the delta rule with training overhead according to the batch size $B$. For our implementation, we set the batch size of the delta rule update to be the full number of vectors contained in the memory matrix, as we find that we can achieve good performance while keeping the delta rule overhead to a minimum.

\subsection*{Neural network and dataset selection}\label{meth:neural_nets}
For our experiments, we utilized publicly available implementations and datasets to ensure reproducibility. The GPT-2-large model~\cite{radford2019language} was obtained via the Huggingface Transformers library, and all language modeling evaluations, as well as \ac{scr}-aware training, were conducted using the WikiText dataset~\cite{wikitext}, specifically \texttt{wikitext-2-raw-v1}, also accessed through Huggingface’s datasets module. For the \ac{dlrm}, we employed the official implementation released by Facebook AI~\cite{Naumov2019}. Training and evaluation of the DLRM were performed on the Criteo Kaggle Display Advertising Challenge dataset, which is publicly distributed by Criteo~\cite{criteo-display-ad-challenge}.

\subsection*{Analog hardware-aware training}\label{meth:hwa_training}
In addition to the SCR-aware training procedure, which makes the model parameters robust to the TPR-induced interference $\etpr$, the memory matrix $\Sp$ must also be made robust to the variability of analog devices. Prior works have shown both in simulation~\cite{Rasch2023,Buechel2025a} and in hardware~\cite{LeGallo2023} that such robustness can be achieved through three main mechanisms during training: clipping, Gaussian noise injection, and output quantization.

To that end, we extend the SCR-aware training procedure to incorporate these elements. Specifically, after constructing $\Sp$ during the third step of the procedure, we apply clipping to the memory matrix using a threshold defined in terms of standard deviations. Gaussian noise is then injected into each element with a standard deviation proportional to its magnitude. After these two steps, the unbinding operation is performed to retrieve the vectors $\hat{\mathbf{V}}$, which are then quantized to emulate the effect of ADC discretization. The rest of the procedure then proceeds as outlined in the respective Methods section.

Adding these analog \ac{ha} elements during training causes the error term that is added to the parameters $\mathbf{P}$, to include both the TPR interference $\etpr$ and hardware-related perturbations such as noise and quantization. As a result, backpropagation converges to parameters that are both robust to the TPR interference and to the analog hardware non-idealities. Importantly, clipping, noise injection, and output quantization are computationally inexpensive steps, and there introduce negligible overhead during training.

In the hardware experiments reported in this work, networks were trained using the hyperparameters listed in Extended Data Table~\ref{ext_tab:hyperparams}. For completeness, Extended Data Fig.~\ref{ext_fig:hwa_ablations} shows the impact of different clipping and embedding quantization settings during training and evaluation under varying levels of injected Gaussian noise.

\subsection*{Hardware Platform}\label{meth:hw_platform}
The hardware platform used in this work is the IBM HERMES Project chip described in Le Gallo et al.~\cite{LeGallo2023}. It comprises 64 compute tiles, each with a 256$\times$256 crossbar array based on \ac{pcm} devices. In this work, the arrays are used to store the memory matrices $\Sp$ with dimensionality $D=256$. A photograph of the chip and the experimental setup is shown in Extended Data Fig.~\ref{ext_fig:hermes_setup}.

To map a matrix $\Sp$ onto the hardware, we employ the one-device programming scheme described in~\cite{LeGallo2023}, resulting in only two physical devices per matrix element and a total of $131,072$ devices per memory array. Specifically for this work, we use chips that contain the Projected \ac{pcm} devices described in~\cite{Syed2023}. The maximum conductance per-device used in our experiments span from $3 \mu S$ to $7 \mu S$, depending on the distribution of the \ac{scr}, such that the on-chip \ac{adc} does not saturate during integration ($I_\mathrm{max} \approx 110 \mu A$)~\cite{KhaddamAljameh2022}.

Key vectors with values in $\{\pm1\}$ are applied to the array using voltage pulses of $0.2\,\mathrm{V}$ and a duration of $130\,\mathrm{ns}$. The minimum pulse duration is limited by the 8-bit input interface of the hardware, which results in a minimum \ac{mvm} time of approximately $127\,\mathrm{ns}$. For that reason, in Section~\ref{sec:outlook}, we include a projection of the performance of hardware that is designed to support bipolar inputs -- commonplace in the literature, due to the popularity of bit-slicing~\cite{Gallo2022,Xue2020,Hung2023}. Array outputs are sensed using current-controlled oscillator-based \acp{adc} and converted on-chip to signed 8-bit integers, which are used as the retrieved vector values in the experiments.

The retrieval accuracy using the analog hardware is depicted for one of the evaluated tasks in Extended Data Fig.~\ref{ext_fig:mvm_accuracy}. The hardware has an effective equivalent precision between 3- and 4-bits (i.e., storing the weights in 4-bits, reading them, and performing the \ac{scr} readout \ac{mvm} in full precision). This falls in line with the demonstrated precision of the analog hardware when used to perform \acp{mvm} for \ac{dnn} inference~\cite{LeGallo2023}.

For the hardware experiments reported in Fig.~\ref{fig:hw_results}, the number of tiled \acp{scr} required for each application was 8 for \ac{dlrm}, 1002 for GPT2-large, and 501 for its $2\times$ compressed version. These experiments were executed by serially utilizing the 64 tiles available on the hardware platform.

\clearpage
\section*{References}
\bibliography{tpr_bibtex}

\section*{Acknowledgements}
\noindent This work was supported by the IBM Research AI Hardware Center. A. Sebastian acknowledges partial funding from the European Research Council (ERC) under the European Union’s Horizon Europe research and innovation programme (grant agreement no. 101201020).
This work was also partially supported by the European Union’s Horizon Europe Research and Innovation Program under Grant Agreement No. 101223271, and by the Swiss State Secretariat for Education, Research and Innovation (SERI) under contract number 25.00443.

\section*{Author Contributions}
\noindent The research was initiated by A. Rahimi and A. Sebastian. A. Vasilopoulos, A. Rahimi, and A. Sebastian jointly developed the research plan. Simulations and hardware experiments were carried out by A. Vasilopoulos and P. Diener. M. Hersche assisted with data analysis. A. Vasilopoulos and A. Sebastian wrote the manuscript with inputs from all the authors.

\section*{Data Availability}
The neural networks in this work were trained and evaluated using the WikiText dataset (\texttt{wikitext-2-raw-v1})~\cite{wikitext}, which is publicly available at \url{https://huggingface.co/datasets/Salesforce/wikitext}, and the Criteo Kaggle Display Advertising Dataset~\cite{criteo-display-ad-challenge}, which is publicly available at \url{https://ailab.criteo.com/ressources/}. 

\section*{Code Availability}
The code used to perform the simulations included in this study can be made available by the corresponding authors upon request.

\section*{Competing Interests}
The authors declare no competing interests.

\clearpage
\section*{Figures}
%-----------------------------------------------------------------------
% Figure 1
%-----------------------------------------------------------------------
\clearpage

%-----------------------------------------------------------------------
% Figure 2
%-----------------------------------------------------------------------
\clearpage

%-----------------------------------------------------------------------
% Figure 3
%-----------------------------------------------------------------------
\clearpage

%-----------------------------------------------------------------------
% Figure 4
%-----------------------------------------------------------------------
\clearpage

\section*{Extended Data}
\setcounter{figure}{0}
\renewcommand{\figurename}{EXTENDED FIG.}
\begin{figure}[ht]
    \centering
    \includegraphics[width=\textwidth]{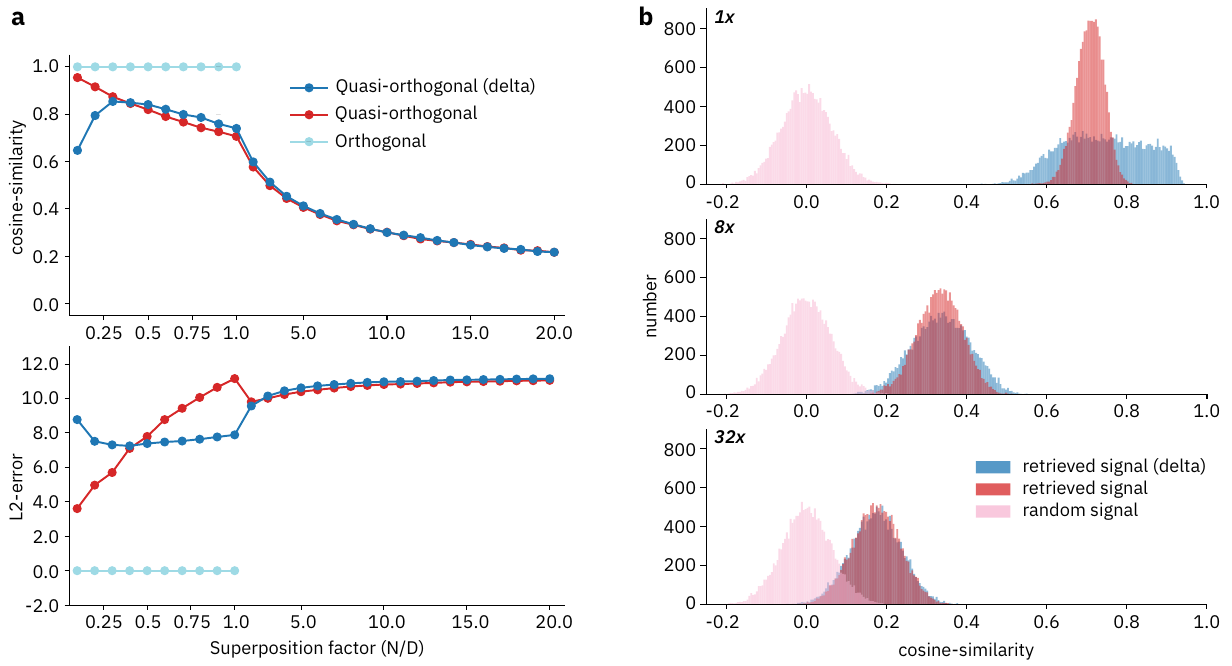}
    \caption{\textbf{SCR properties with increasing superposition.}
    \textbf{a,} Retrieval accuracy of vectors stored in the superposed composite representation $\Sp$ as a function of the number of stored vectors relative to dimensionality ($N/D$). Random value vectors $\vi$ are drawn from a Gaussian distribution $\mathcal{N}(\boldsymbol{\mu},\boldsymbol{\sigma})$. When orthogonal keys are used, retrieval is exact, and both cosine similarity and normalized L2-error indicate perfect reconstruction. However, orthogonality restricts the number of stored vectors to $N\le D$. When quasi-orthogonal keys drawn from a Rademacher distribution are used, cosine similarity degrades gradually as $N$ increases beyond $D$, while the L2-error grows because the magnitude of the retrieved vector deviates from the original value. This behaviour reflects the preservation of vector direction despite increasing interference from other stored vectors. The two superposition functions considered -- simple addition and the modified delta update rule -- exhibit similar behavior for superposition factors larger than one ($N>D$), while small differences appear for low superposition levels, as the delta update rule is not optimized for that regime.
    \textbf{b,} Distribution of cosine similarity between the retrieved vectors $\{\vhati\}$ and their corresponding stored vectors $\{\vi\}$ for quasi-orthogonal keys. For $N=D$, retrieved signals remain well separated from random vectors drawn from the same distribution but not stored in the representation. As more vectors are superposed ($N>D$), the distributions gradually overlap with random noise, indicating increasing interference. Nevertheless, useful signal separation remains visible even for substantial superposition factors (for example, $N=8D$ and $N=32D$), illustrating the graceful degradation of retrieval quality.
    Retrieval accuracy is evaluated using two metrics: the normalized L2-error, $E_{\mathrm{L2}} = \|\vhati-\vi\|_2 / \|\vi\|_2$, which measures the relative magnitude difference between retrieved and stored vectors, and cosine similarity, $\cos\theta = \vhati^\top \vi / (\|\vhati\|_2 \|\vi\|_2)$, which measures their angular alignment. Points in panel \textbf{a} correspond to averages over a batch of vectors, while panel \textbf{b} shows the distribution of cosine similarity across individual vectors.
    }
    \label{ext_fig:tpr_graceful_degradation}
\end{figure}
\clearpage

\begin{figure}[ht]
    \centering
    \includegraphics[width=0.8\textwidth]{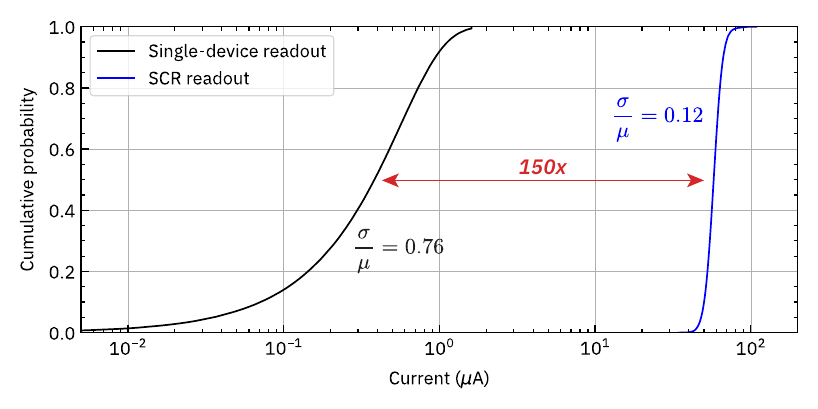}
    \caption{\textbf{Readout current comparison} between single-device reads and SCR reads on the GPT2 embeddings. The collective readout mechanism provides $150\times$ higher median current and a more packed distribution overall compared to the single-device reads. Differential readout is assumed for \ac{scr}, as described in Fig.~\ref{fig:hw_results}. A maximum device conductance of $\sim7 \mu S$ was used during the programming of the arrays on the hardware platform used in this work~\cite{LeGallo2023}.}
    \label{ext_fig:current}
\end{figure}
\clearpage

\begin{figure}[ht]
    \centering
    \includegraphics[width=0.5\textwidth]{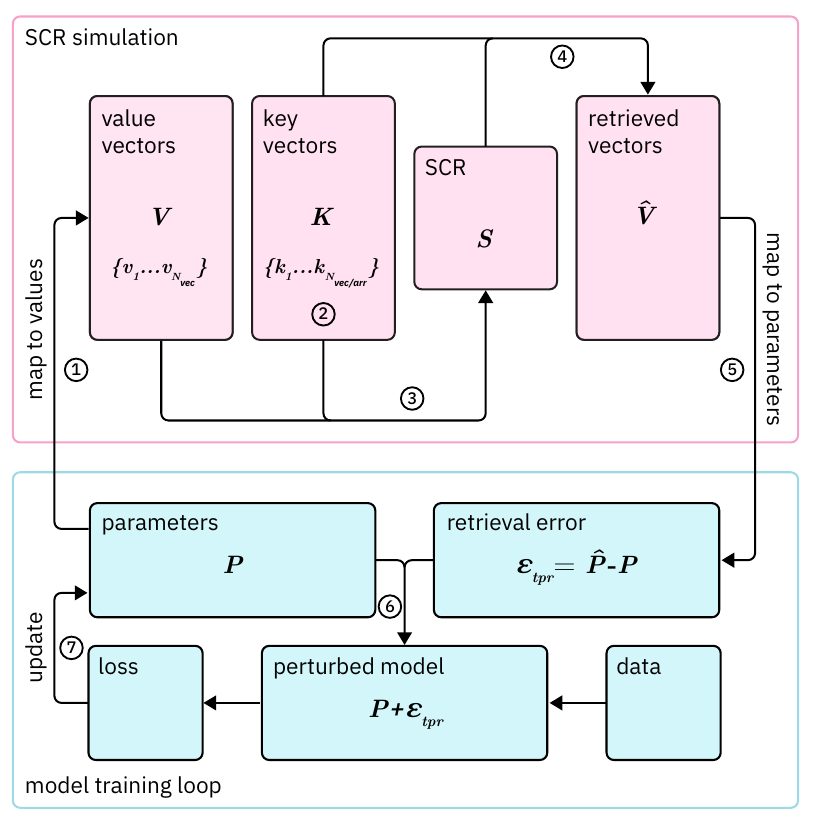}
    \caption{\textbf{SCR-aware training procedure}. This figure describes pictorially the SCR-aware training procedure described in the Methods section. Steps are numbered according to the text to guide the reader.}
    \label{ext_fig:training}
\end{figure}
\clearpage

\begin{figure}[ht]
    \centering
    \includegraphics[width=0.75\textwidth]{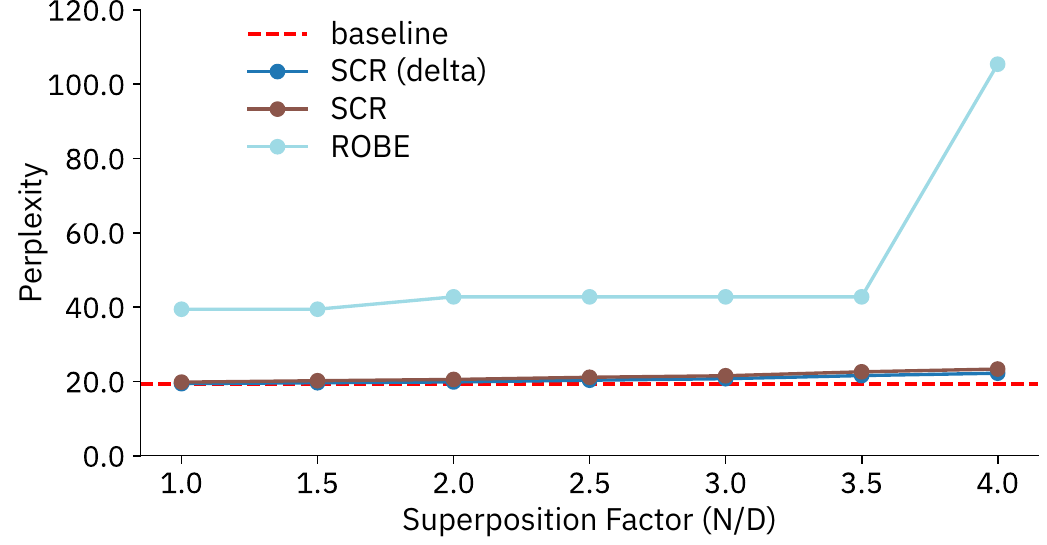}
    \caption{\textbf{Generalization of SCR across tasks.} The compression behaviour of the proposed method generalizes across both \ac{dlrm} and GPT2-large embedding tables. In contrast, the state-of-the-art \ac{dlrm} compression method ROBE fails to generalize to GPT2-large embeddings, even without compression. This difference arises because the \ac{scr} representation is distribution-agnostic: the memory state is constructed directly from the stored vectors through key binding, without assumptions about their underlying distribution.}
    \label{ext_fig:gpt_robe}
\end{figure}
\clearpage

\begin{figure}[ht]
    \centering
    \includegraphics[width=\textwidth]{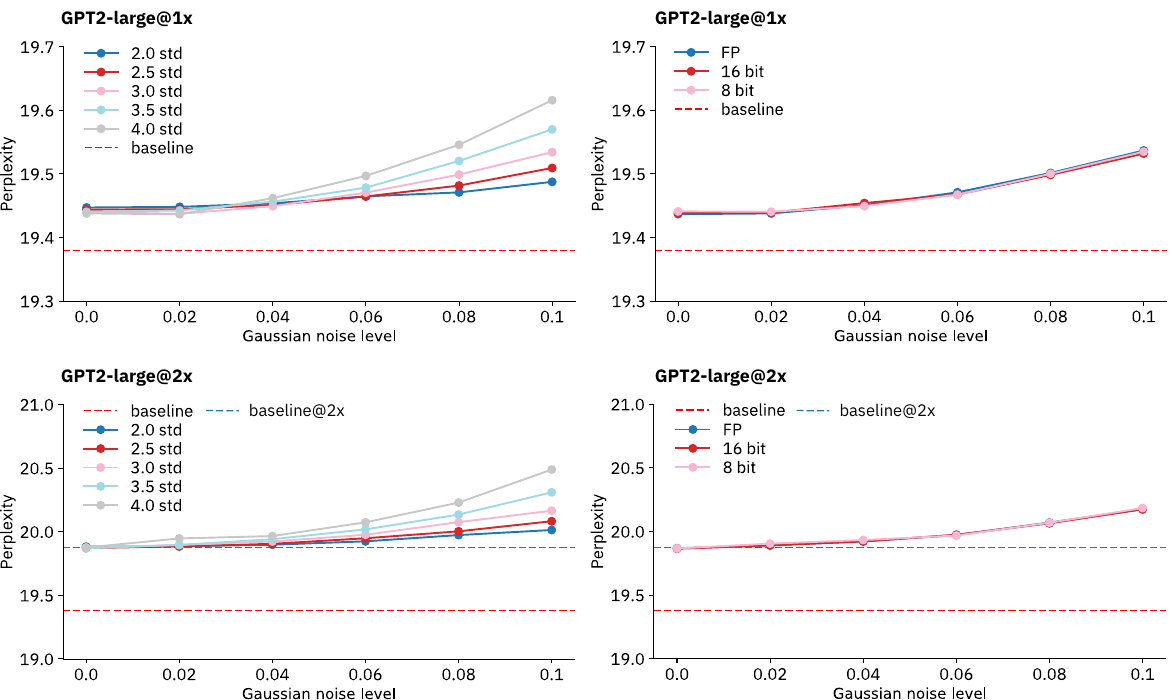}
    \caption{\textbf{Ablation study of hardware-aware training hyperparameters.} We observe that the clipping threshold directly affects robustness to analog noise. Stricter clipping (e.g., 2 standard deviations) produces models that remain more stable under the same noise level than models trained with weaker clipping thresholds. By contrast, output quantization down to 8 bits has little impact on the precision of the models, performing comparably to the non-quantized case.}
    \label{ext_fig:hwa_ablations}
\end{figure}
\clearpage

\begin{figure}[ht]
    \centering
    \includegraphics[width=\textwidth]{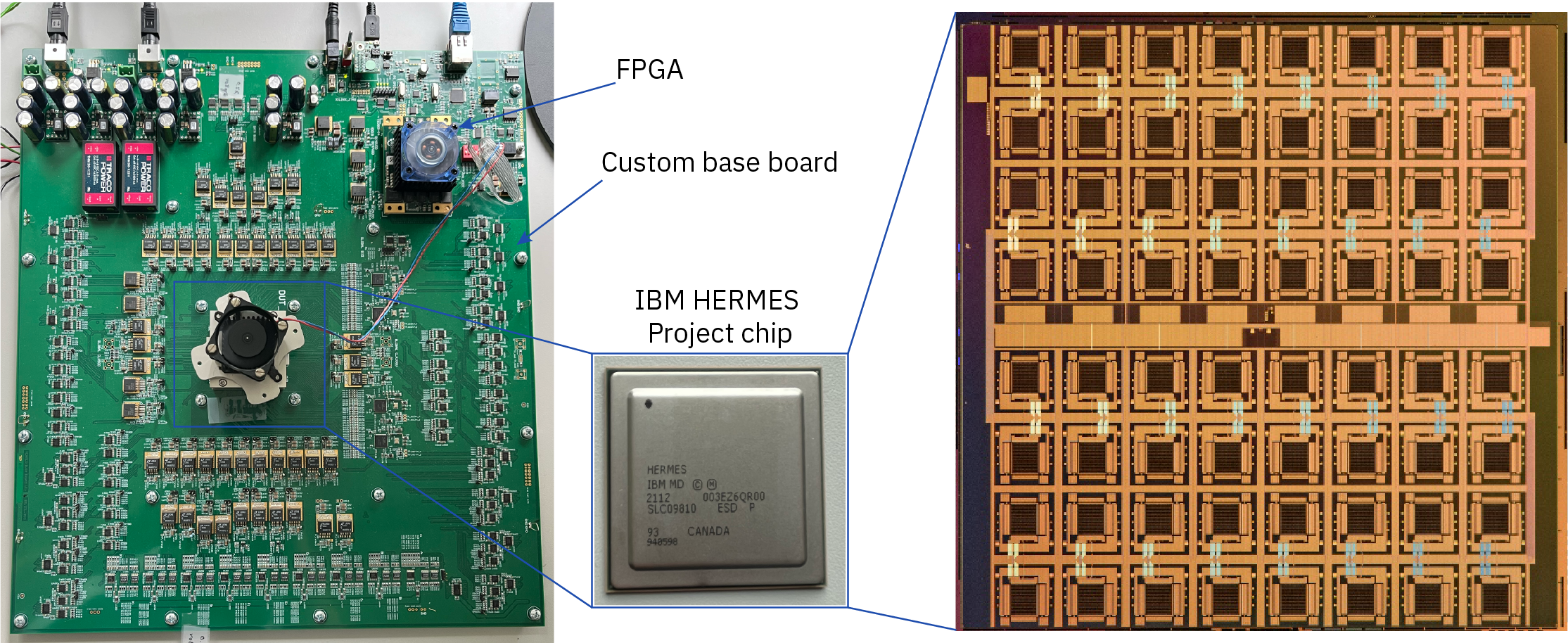}
    \caption{\textbf{The IBM HERMES Project chip hardware platform~\cite{LeGallo2023}}. The experimental platform consists of three main components. A custom baseboard provides power delivery, clock distribution, and discrete electronics required to operate the chip, as well as interfaces to a host PC. Communication with the host is performed through JTAG for \ac{fpga} programming and control, and through Ethernet for data transfer. Second, an off-the-shelf Xilinx Trenz TE0808-04 module with a Xilinx Zynq UltraScale+ \ac{fpga} controls the baseboard electronics and provides the interface between the host PC and the chip through its serial input/output interface. Finally, the packaged IBM HERMES Project chip is mounted on the baseboard socket and accessed through the \ac{fpga}-controlled interface.}
    \label{ext_fig:hermes_setup}
\end{figure}
\clearpage

\begin{figure}[ht]
    \centering
    \includegraphics[width=\textwidth]{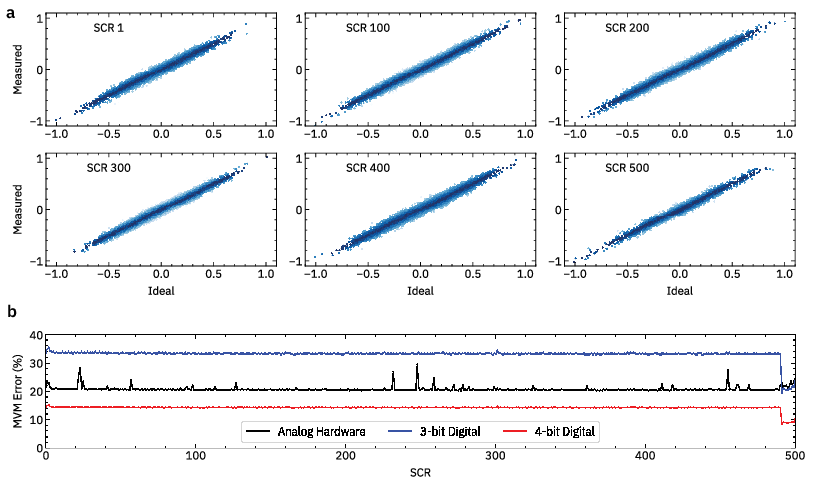}
    \caption{\textbf{SCR retrieval MVM accuracy} on the GPT2 embeddings at $2\times$ superposition. The analog hardware performs has an effective precision 3- and 4-bit digital equivalent, closer on average to the 4-bit baseline. This demonstrates effective use of the analog states of the devices.}
    \label{ext_fig:mvm_accuracy}
\end{figure}
\clearpage

\renewcommand{\tablename}{EXTENDED DATA TABLE}
\setcounter{table}{0}
\begin{table}
\centering
\begin{tabularx}{0.75\textwidth}{|c|Y|Y|Y|}
\hline
\textbf{Hyperparameter} & \textbf{DLRM} & \textbf{GPT2-large@1x} & \textbf{GPT2-large@2x} \\
\hline
\textbf{Learning rate} & 1.0 & 1e-5 & 1e-5 \\
\textbf{Batch size} & 512 & 32 & 32 \\
\textbf{Delta rule $\beta$} & 0.5 & 0.5 & 0.5 \\
\textbf{Gaussian noise (\%)} & 8 & 10 & 10 \\
\textbf{$\Sp$ clipping (std)} & 3 & 3 & 3 \\
\textbf{Output quantization (bits)} & 8 & 8 & 8 \\
\hline
\end{tabularx}
\caption{\textbf{Training hyperparameters for all experiments.} The fields learning rate and batch size are used in all experiments, while the rest of the fields are only applicable to the appropriate experiments. Specifically, the delta rule $\beta$ only applies when the delta update rule is used, while the Gaussian noise, the clipping, and the output quantization are only used in the hardware-aware training and simulations.}
\label{ext_tab:hyperparams}
\end{table}
\clearpage

\begin{table}
\centering
\begin{tabularx}{\textwidth}{|Y|Y|Y|Y||Y|Y|}
\hline
\multicolumn{4}{|c||}{\textbf{Hardware parameters}} & \multicolumn{2}{c|}{\textbf{Memory performance}} \\
\hline
\textbf{Array Size $\mathbf{(D\times D)}$} & \textbf{Energy per MAC (fJ)} & \textbf{Periphery Overhead ($\times E_{array}$)} & \textbf{Read time (ns)} & \textbf{Energy per effective bit read (pJ)} & \textbf{Bandwidth (GB/s)} \\
\hline
512$\times$512 & 0.5 & 10 & 10 & 0.704 & 25.6 \\
 & & & 5 & 0.704 & 51.2 \\
 & & 5 & 10 & 0.384 & 25.6 \\
 & & & 5 & 0.384 & 51.2 \\
 & 0.05 & 10 & 10 & 0.0704 & 25.6 \\
 & & & 5 & 0.0704 & 51.2 \\
 & & 5 & 10 & 0.0384 & 25.6 \\
 & & & 5 & 0.0384 & 51.2 \\
\hline
1024$\times$1024 & 0.05 & 5 & 5 & 0.0768 & 102.4 \\
\hline
2048$\times$2048 & 0.05 & 5 & 5 & 0.1536 & 204.8 \\
\hline
4096$\times$4096 & 0.05 & 5 & 5 & 0.3072 & 409.6 \\
\hline
\end{tabularx}
\caption{\textbf{Analog memory array-level performance and energy projections.} We use the following formulas to calculate the \textit{Energy per-bit} ($E_{per-bit}$) and \textit{Bandwidth} ($BW$) according to the hardware parameters: $E_{per-bit}=\frac{D^2 * E_{MAC} * (1+Overhead)}{B_{effective}*D}*10^{-3}$ and $BW=\frac{B_{effective}}{8}*\frac{D}{t_{read}}$. We assume an effective precision for the retrieval $B_{effective}=4$, in line with effective weights storage precision in published \ac{aimc} hardware~\cite{Vasilopoulos2023,Buechel2023}. Baseline parameters, such as array size of $512\times512$~\cite{Ambrogio2023}, energy per-MAC of 0.5 fJ~\cite{LeGallo2023,Vasilopoulos2023}, ~10$\times$ periphery overhead~\cite{LeGallo2023} and read time~\cite{Xue2020} are taken from published \ac{nvm}-based \ac{aimc} hardware, while the table also explores various projections to future improved hardware parameters, i.e., larger array size, improved energy per MAC due to more resistive devices, more energy efficient periphery, and faster read time.
}
\label{ext_tab:performance_proj}
\end{table}
\end{document}